\documentclass[preprint2]{aastex}
\usepackage{spr-astr-addons}
\usepackage{url}
\usepackage{amssymb}

\begin{document}

\title{An Optical-UV Survey of the North Celestial Cap}
\shorttitle{Optical-UV Survey of NCC}
\shortauthors{Gorbikov \& Brosch}

\author{Evgeny Gorbikov\altaffilmark{1}} \and \author{Noah Brosch\altaffilmark{1}}
\email{evgenyg1@post.tau.ac.il}
\altaffiltext{1}{The Wise Observatory and the Raymond and Beverly Sackler School of Physics and Astronomy, the Faculty of Exact Sciences, Tel Aviv University, Tel Aviv 69978, Israel}

\begin{abstract}
We present preliminary results of an optical-UV survey of the North Celestial Cap (NCCS) based on $\sim$5\% areal coverage. The NCCS will provide good photometric and astrometric data for the North Celestial Cap region ($80^{\circ}\leq\delta\leq90^{\circ}$). This region, at galactic latitudes from $17^{\circ}\lesssim b\lesssim37^{\circ}$, is poorly covered by modern CCD-based surveys. The expected number of detected objects in NCCS is $\sim$1,500,000. We discuss issues of galactic structure, extinction, and the galaxy clustering in the colour-colour diagrams.

\end{abstract}
\vspace{-8pt}
\vspace{-8pt}
\section{Introduction}
\vspace{-8pt}
From the dawn of humankind people were interested in studying the surrounding world. In ancient times astronomy, and sky mapping in particular, had not only a world-description function, but had also practical purposes. For example, sky maps were used for navigation and orientation as well as for predicting celestial phenomena.

Two inventions contributed significantly to sky mapping: the telescope and the permanent photography. Photographic catalogs, produced from the 1890s to the 2000s, play an important role even today. The first high-quality digital all-sky survey, the \textbf{Digitized Sky Survey} (DSS) and its extension DSS-II, were produced by scanning photographic survey plates (POSS-I, POSS-II, ESO/SERC) with specific photometric calibrations. The POSS-I, POSS-II and ESO/SERC photographic surveys provided high-precision astrometry used in catalogs such as \textbf{USNO-A1.0}, \textbf{USNO-A2.0}, \textbf{USNO-B1.0}, etc. USNO-A and USNO-B are all-sky high-precision astrometric catalogs including also photometric data from the DSS and DSS-II. In many cases, the USNO DSS-based catalogs are the only optical high-precision data available for a particular sky area. 

However, photographic catalogs suffer from significant photometric and astrometric systematic and statistical errors due to shortcomings of the photographic emulsion, such as low sensitivity, limited dynamic range and non-linearity (\citealt{MON03}). These were solved with the introduction of the charge-coupled device (CCD) to astronomy. CCDs eliminated not only the photographic emulsion shortcomings, but solved also other problems, such as providing an effective raw data storage, yielding a short delay between the raw data collection and the final data extraction, and being a reusable photosensitive element. All modern optical sky surveys are CCD-based.

The modern optical sky survey in general use is the \textbf{Sloan Digital Sky Survey} (SDSS), which covers about 10,000 deg$^2$ of the sky and provides high-precision photometric and astrometric data in five Sloan bands (\citealt{ABA09}).

An important extension to the optical catalogs are UV observations. One of the prominent available UV instruments is the \textbf{Galaxy Evolution Explorer} (GALEX), performing both imaging and low-resolution spectroscopy in two bands: near-UV (NUV). GALEX conducts several pioneering UV sky surveys aimed primarily to understand galaxy evolution, and its publicly-available dataset covers by now about 3/4 of sky (\citealt{MOR07}).
\vspace{-8pt}
\section{The North Celestial Cap Survey}
\vspace{-8pt}
\subsection{Overview}
\begin{figure}[ht!]
\begin{center}
\begin{tabular}{cc}
\includegraphics[angle=0,width=0.215\textwidth]{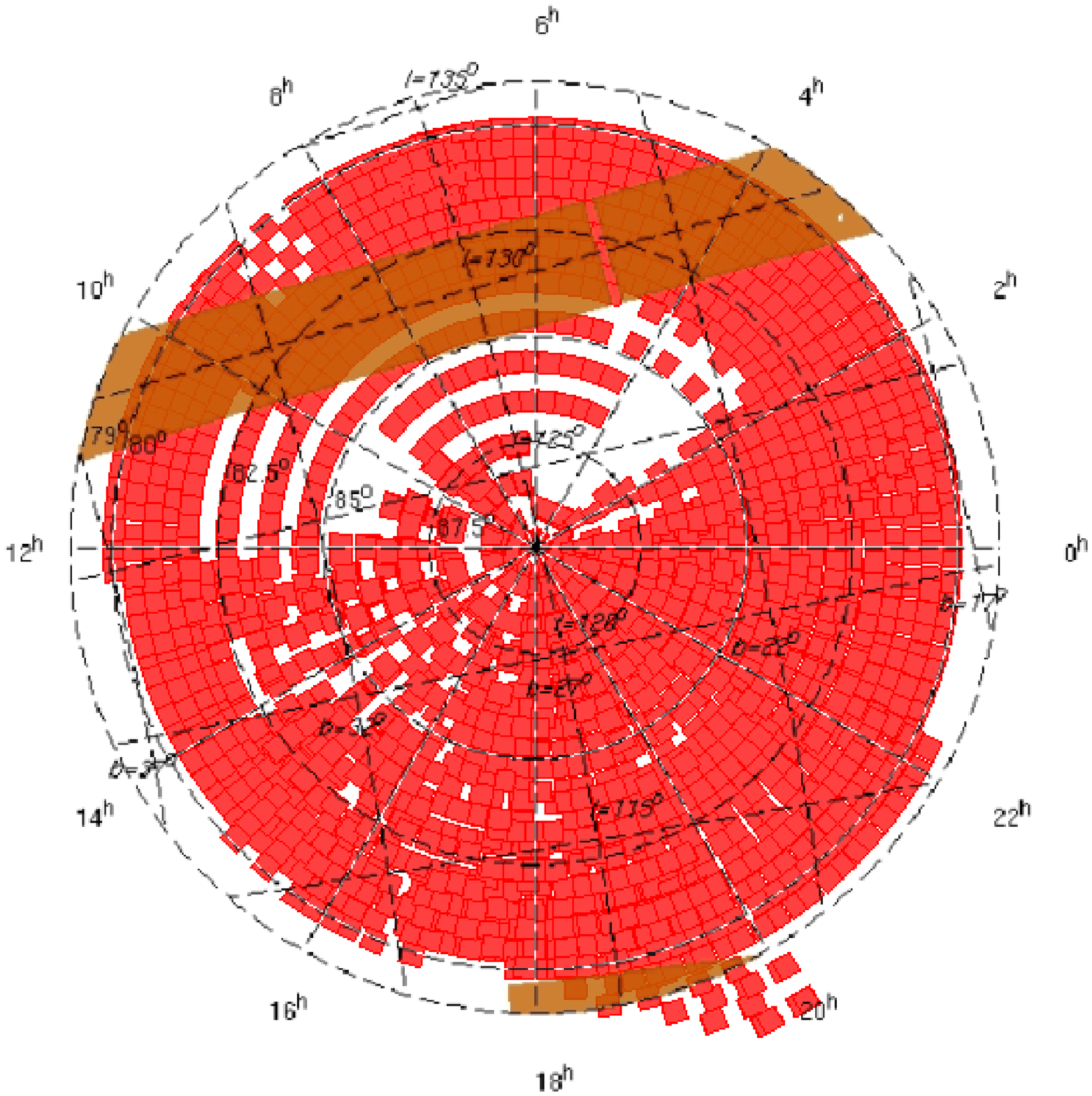} &
\includegraphics[angle=0,width=0.215\textwidth]{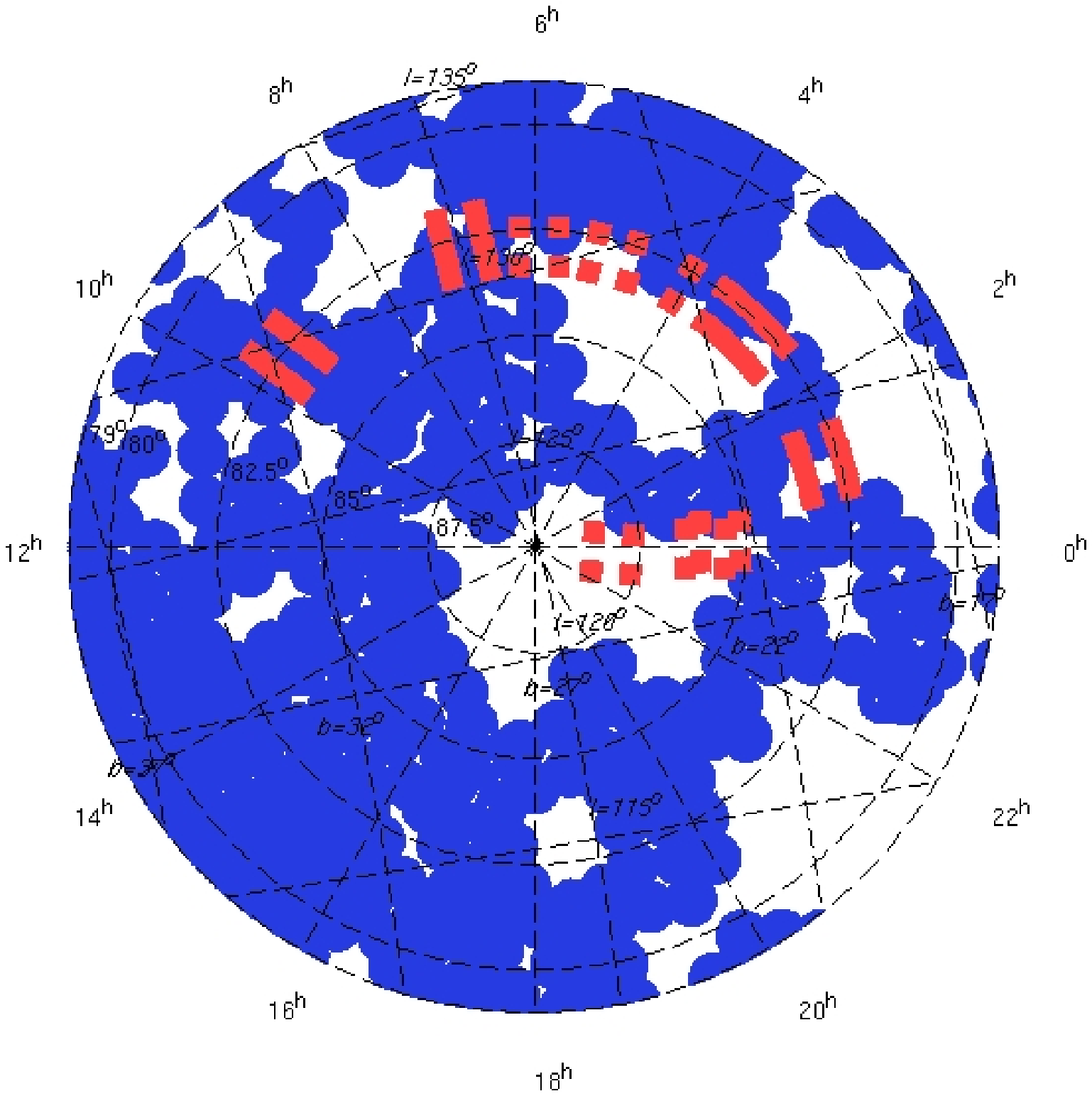} \\
\textit{a} & \textit{b}
\end{tabular}
\end{center}
\caption{\textit{Panel a}: NCC coverage by the survey in R and I bands as for July, 27, 2010 (red) and by SEGUE-SDSS (brown). \textit{Panel b}: NCC coverage by GALEX (blue) and the selected area (red)
   \label{Fig:coverage}}
\end{figure}

\vspace{-8pt}
The \textbf{North Celestial Cap Survey} (NCCS) conducted at Wise Observatory, Israel, will obtain good photometric and astrometric data for the North Celestial Cap (NCC) region ($80^{\circ}\leq\delta\leq90^{\circ}$). A significant portion of the data collection is performed at Wise Observatory with the 1-meter Ritchey-Chr\'{e}tien telescope and the Large Area Imager for the Wise Observatory (LAIWO) camera (\citealt{GOR10}). The R and I band imaging is $\sim$90\% complete and will be completed in about six months. The current coverage is shown in Fig.\ref{Fig:coverage}\textit{a}. Based on preliminary results, the final optical catalog will include $\gtrsim$1,500,000 distinct objects. The photometric accuracy is $\Delta $m~$ \lesssim 0^m.1$ for point sources brighter than R~$=$~20$^m$.2 and I~$=$~19$^m$.1, and the catalog will be complete to these magnitudes. The astrometric solution is derived using the \textbf{third US Naval Observatory CCD Astrograph Catalog} (UCAC3, \citealt{UCAC3}) and the demonstrated accuracy is $\sim$0$''$.1--0$''$.3 for the both $\alpha$ and $\delta$. The NCCS point-extended source separation (PESS) was checked against the SDSS and the PESS accuracy was found to be $>$90\% accurate (\citealt{GOR10}).

Another important part of the project is collecting data also in Sloan g' band to extend the wavelength coverage and to fill the gap between the red optical filters and UV filters. We expect to add Sloan g' band observations with the 48-inch Schmidt telescope of Palomar Observatory equipped with the \textbf{Palomar Transient Factory} (PTF, \citealt{LAW09}) camera.

The third part of the survey, the GALEX data (GR4/GR5), was extracted from the GALEX web site. These originate from the AIS survey, since yet there is no MIS coverage in the NCC region. The AIS coverage in the NCC region is $\sim$50\% and is shown in Fig.\ref{Fig:coverage}\textit{b}. There are $\sim$10$\times$ more sources detected in NUV than in FUV.
\vspace{-8pt}
\subsection{Motivation and Purposes}
\vspace{-8pt}
The NCCS project was originally planned to support the \textbf{Tel Aviv University UV Experiment} (TAUVEX) mission by extending the wavelength base where each source was measured from the UV to the optical (\citealt{GOR10}). In the last planned version, TAUVEX was expected to observe objects by scanning around the North and the South Celestial Poles (\citealt{ALM07}). These celestial cap regions are poorly covered by modern optical surveys, but the northern sky patch can be easily observed from the Wise Observatory.

The DSS-based all-sky photographic catalogs, such as USNO-B1.0, do cover the NCC region. However, they suffer from significant photometric statistical and systematic errors, as mentioned above. SDSS is deeper and its photometry is more accurate than that of USNO-B1.0, but it covers only a tiny fraction of the NCC region (Fig.\ref{Fig:coverage}\textit{a}). The NCCS was planned to survey the NCC region and provide photometry better by $\sim$4$\times$ relative to USNO-B1.0 for stars at the limiting magnitude and even more accurate for brighter objects. Although NCCS is shallower than SDSS by $\sim$1.5--2.0 mag, it will fully cover the entire NCC region.

Since the TAUVEX UV data is not available, GALEX data (GR4/GR5) was used for the UV magnitudes. Only $\sim$50\% of the NCC region are covered by the All Sky Survey (AIS) of GALEX in the current data release, as shown in Fig.\ref{Fig:coverage}\textit{b}, although more coverage is expected to become available at the next data release. Using GALEX data is a good option, since its data are sufficiently deep and the NCC coverage by GALEX is fairly complete.

The NCCS will fill the time gap produced between the completion of the SDSS project and the beginning of next-generation major sky surveys, such as Pan-STARRS and LSST. The survey location is of high interest since it is located at intermediate galactic latitudes ($17^{\circ}\leq b \leq 37^{\circ}$) where both the Milky Way (MW) stellar and interstellar dust structures, and extragalactic objects, can be studied. The primary scientific purposes of the survey are (a) a study of the MW stellar structure, and (b) of galactic extinction at intermediate latitudes, (c) the detection of fast-moving objects, (d) the identification of white dwarf candidates, QSOs and AGNs, and (e) the study of galaxy bimodality in a sky region not previously surveyed for this. 
\vspace{-8pt}
\section{Preliminary Results}
\vspace{-8pt}
\subsection{Selected Area}
\vspace{-8pt}
To demonstrate the expected yield from the NCCS we present below preliminary results for a 14 deg$^2$ region imaged during a single night  (January, 7, 2010) and shown in Fig.\ref{Fig:coverage}\textit{b}. The objects were photometrically calibrated relative to \cite{LAN09} standards observed in the same night. More than 100,000 sources were detected in each band by the pipeline. Sources detected in two bands were matched and $\sim$77,000 of them were found to be brighter than R~$=$~20$^m$.2 and I~$=$~19$^m$.1. The pipeline classified $\sim$55\% of the objects as extended.

The optical detections were matched with GALEX sources. The GALEX coverage in the selected region is $\sim$50\%, similar to the entire survey region. 12,788 optical objects have a NUV counterpart and 1,433 have a FUV counterpart.
\vspace{-8pt}
\subsection{Galactic Extinction}
\vspace{-8pt}
To measure the galactic extinction in the test area we use the \cite{WOL23} diagram comparing the cumulative star count distributions in an examined and in a reference region, normalized to the same solid angle, and the 'Bull's eye' method (\citealt{GOR10ext}) comparing the distributions in two concentric areas, taking the inner region as the examined one and the outer region as the reference. These methods can only determine the relative extinction between two regions, thus we aim to detect only localized, small-scale dust features. 

The selected region was scanned with an inner radius of 1/4 deg and an outer radius of 1/2 deg in each of the available bands, excluding the FUV band, since GALEX did not detect sufficient FUV sources in this region. Only objects defined by the NCCS pipeline as point sources were used for this study. The results were compared with the \cite{SCH98} $E_{B-V}$ values transformed to extinction values using the \cite{SCH98} relations for $A_R, A_I$ and the \cite{SEI05ext} relations for $A_{NUV}$ and $A_{FUV}$. The relative extinction in four bands at each location was calculated for the comparison.

The median reddening in the selected region from \cite{SCH98} is $\tilde{E}_{B-V} \sim0^m.16$ with a maximal value of 0.51 mag, consistent with the reddening for the entire survey region. In general, our results agree with \citeauthor{SCH98} Most extinction values are within 1$\sigma$ of those predicted by \citeauthor{SCH98}, although there seem to be some exceptions. The extinction law for the selected region cannot be derived using the currently available data, however, the extinction law seems to be closer to the normal MW law than to grey extinction. 
\vspace{-8pt}
\subsection{Stellar Structure}
\vspace{-8pt}
The survey region is located at intermediate galactic latitudes, where, in principle, both thick disk and halo stellar populations can be observed. 

\begin{figure}[ht!]
\begin{center}
\begin{tabular}{cc}
\includegraphics[angle=0,width=0.215\textwidth]{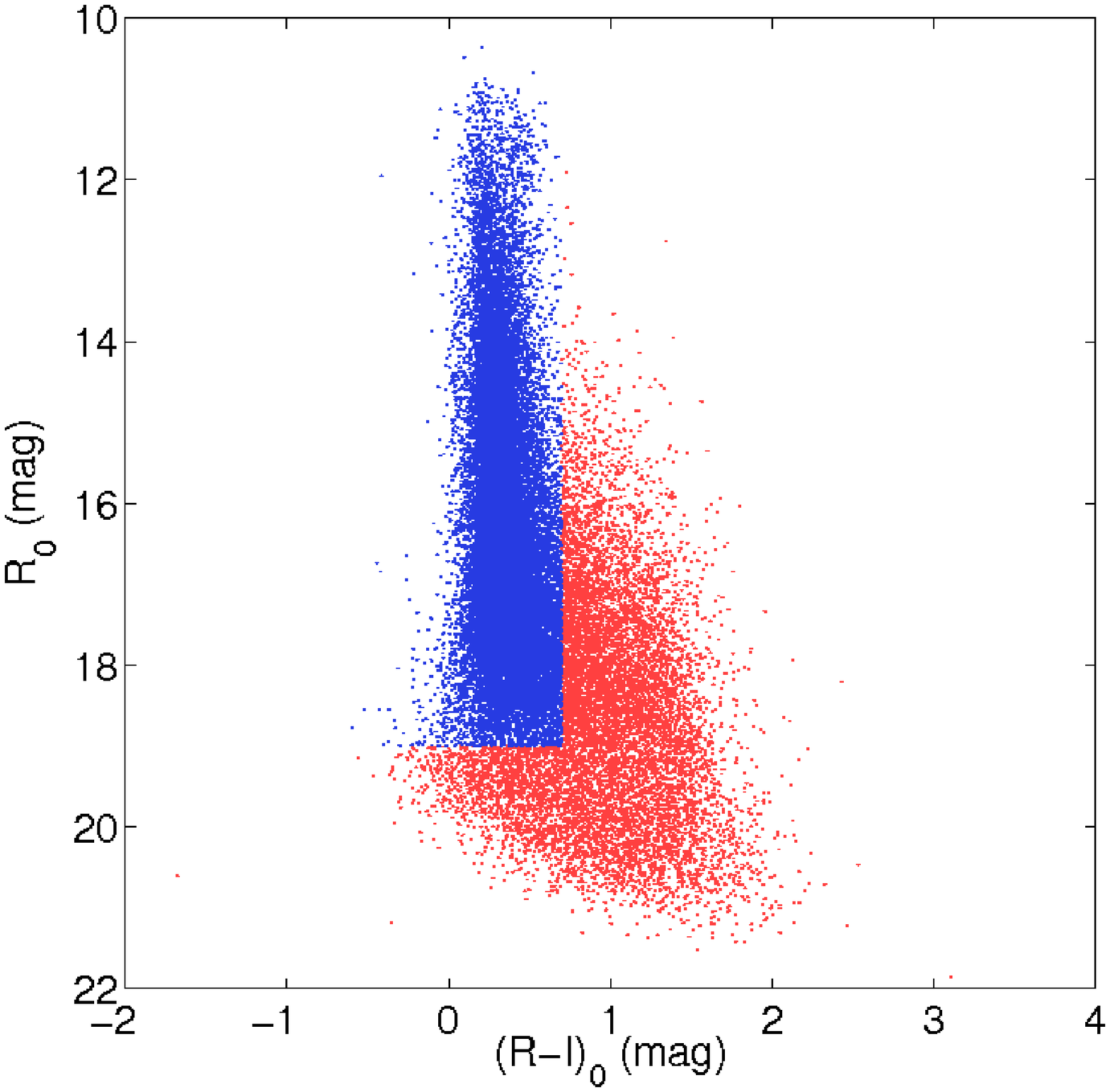} &
\includegraphics[angle=0,width=0.215\textwidth]{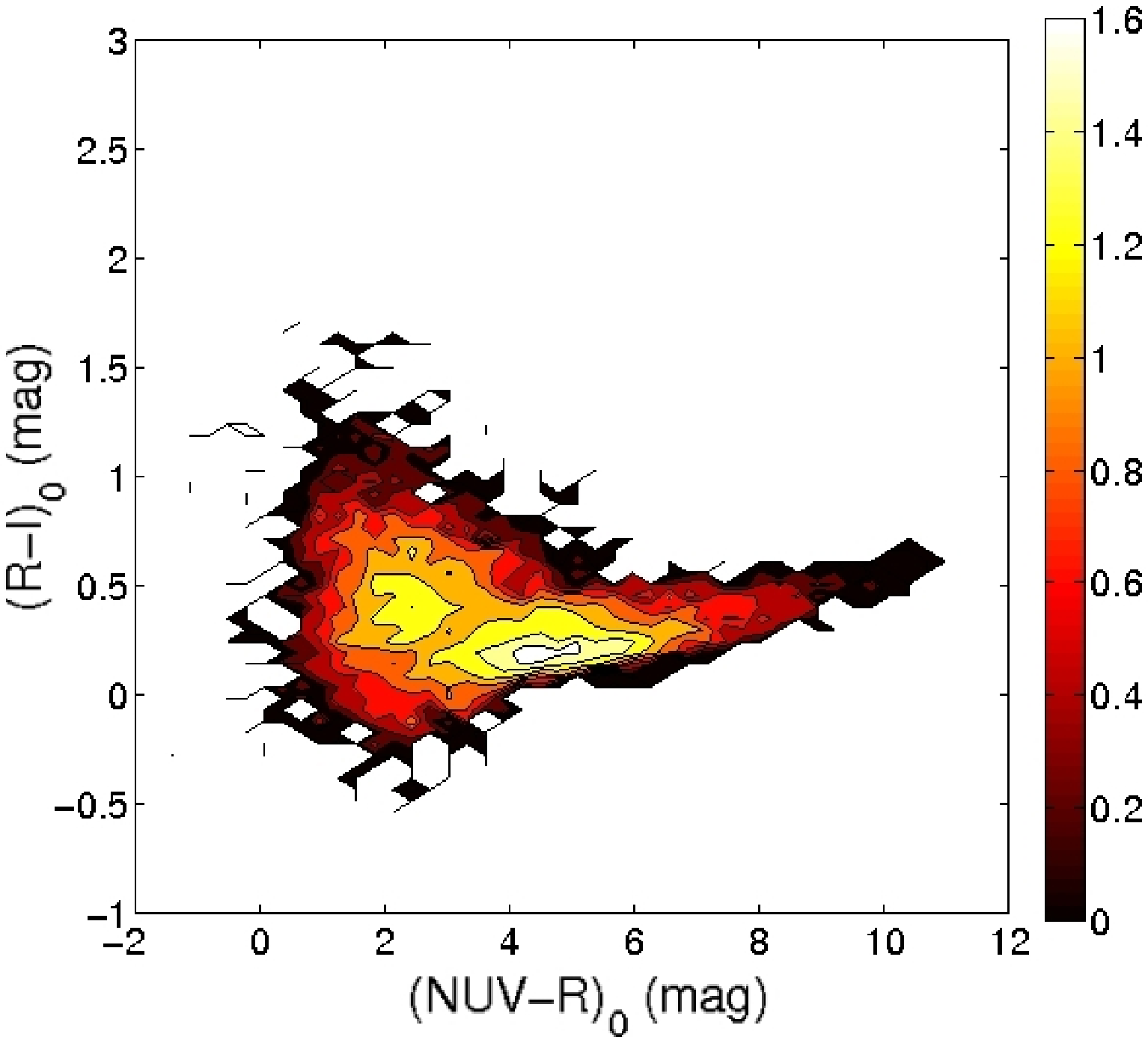} \\
\textit{a} & \textit{b}
\end{tabular}
\end{center}
\caption{\textit{Panel a}: The colour-apparent magnitude stellar distribution of the selected region. \textit{Panel b}: Colour-colour density diagram of the galaxies in the selected region. The intensity is a galaxy count log per squared colour bin
   \label{Fig:Hess}}
\end{figure}

Objects in the selected region defined as point sources by the NCCS pipeline were de-reddened using the \cite{SCH98} reddening map and assuming foreground dust, as done by \cite{dJO10}. The colour-apparent magnitude diagram  for the stellar content of the selected region are shown in Fig.\ref{Fig:Hess}\textit{a}. We divided the stellar distribution into two ares - the blue and the red one, as shown in Fig.\ref{Fig:Hess}\textit{a}. The distances for the stars in the blue and the red regions were calculated assuming main-sequence stars; this is valid for $\sim$90\% of the stars \citep{MEN79}, and using the \cite{COX00} HR diagram. 

The stellar distances for the blue region concentrate at $\sim$2.4 kpc, consistent with a thick disk population, while the stellar distance distribution for the red region is more extended and concentrates at $\sim$3.9 kpc, consistent with a halo population. It is still difficult to separate the two populations, since they mutually contaminate and usually one colour is not sufficient to determine unequivocally the object distance. For instance, \cite{GOR10ext} used two optical colours for a more accurate distance estimation. 
\vspace{-8pt}
\subsection{Galaxy Colour Bimodality}
\vspace{-8pt}
A clear galaxy bimodality in colour-colour space was first demonstrated by \cite{SEI05} using GALEX NUV-based color-color plots. They also found that different types of galaxies (ellipticals, spirals, starburst galaxies, Seyferts and LINERs) reside in different locations on this diagram.

All the galaxies from our preliminary sample were selected and de-reddened for MW dust assuming \cite{SCH98} foreground reddening. Although we use different colours, the results are similar to those of \cite{SEI05} and a clear bimodality in galaxy colours is present, see Fig.\ref{Fig:Hess}\textit{b}. 
\vspace{-8pt}
\subsection{Exotic Objects}
\vspace{-8pt}
While mining the preliminary dataset a few exotic objects of high interest were identified. One such object is a star located at $(\alpha,~\delta)=$ $(03^{h}27^{m}11^{s}.7$, $82^{\circ}15'47''.9)$. 
This star was noted for its blue colours: R = 17$^m$.44, I = 17$^m$.44, NUV = 18$^m$.09, FUV = 18$^m$.71, corresponding to an A-type star. The star is located in a relatively high-extinction region: E$_{B-V}$ = 0$^m$.29 (\citealt{SCH98}), thus it may be of an even earlier spectral type. 

\begin{figure}[ht!]
\begin{center}
\begin{tabular}{ccc}
\includegraphics[angle=0,width=0.135\textwidth]{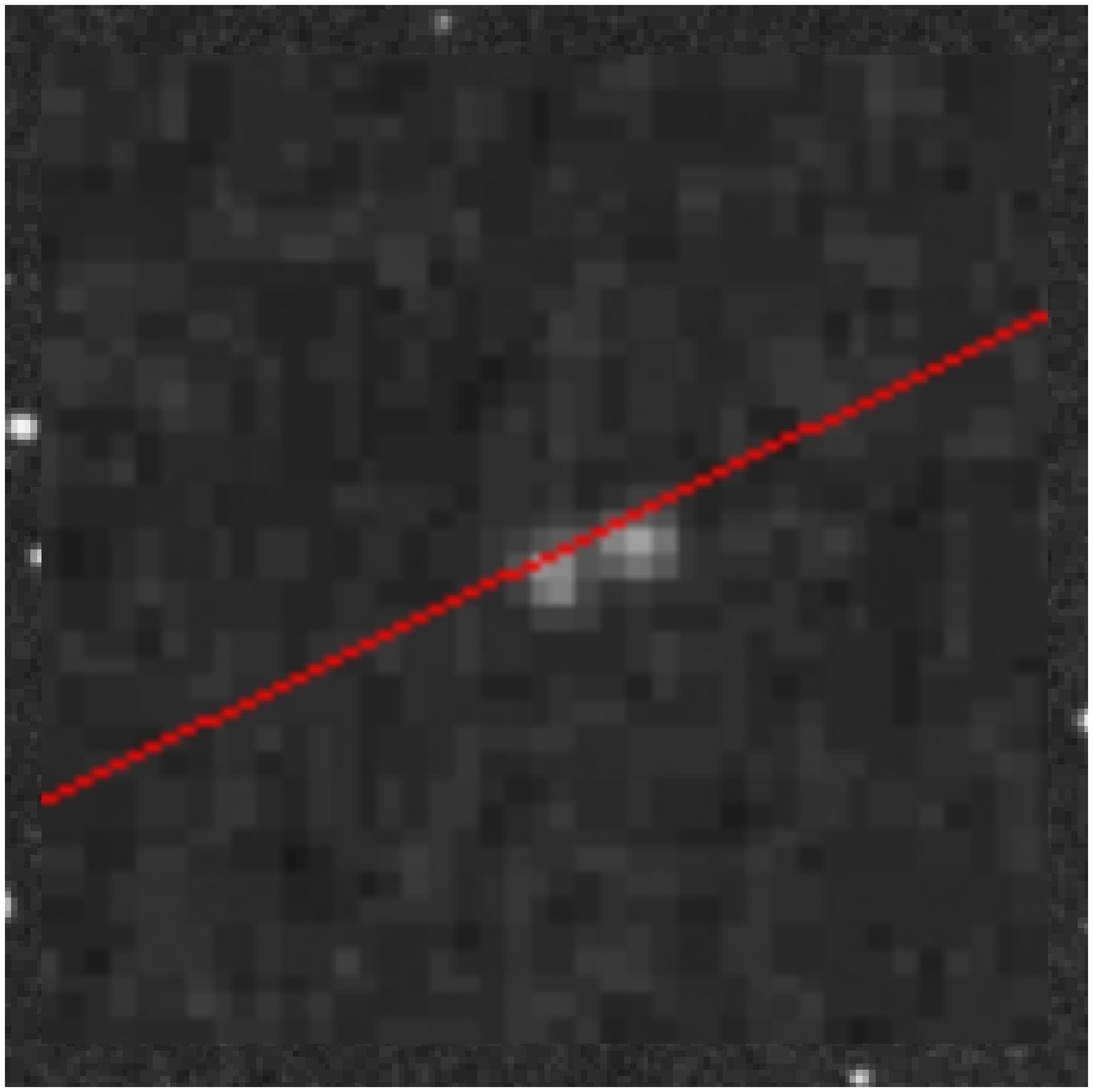} &
\includegraphics[angle=0,width=0.135\textwidth]{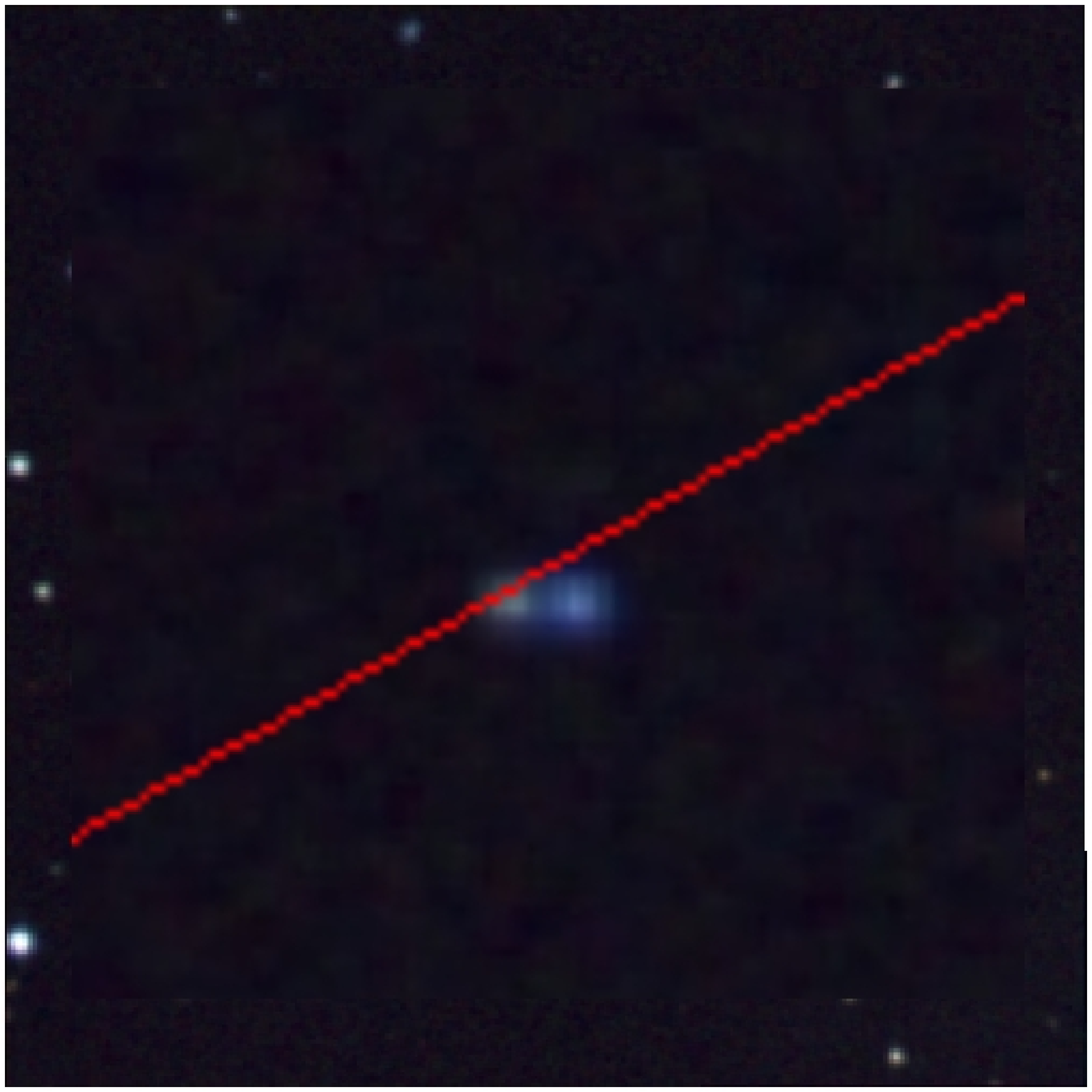} &
\includegraphics[angle=0,width=0.135\textwidth]{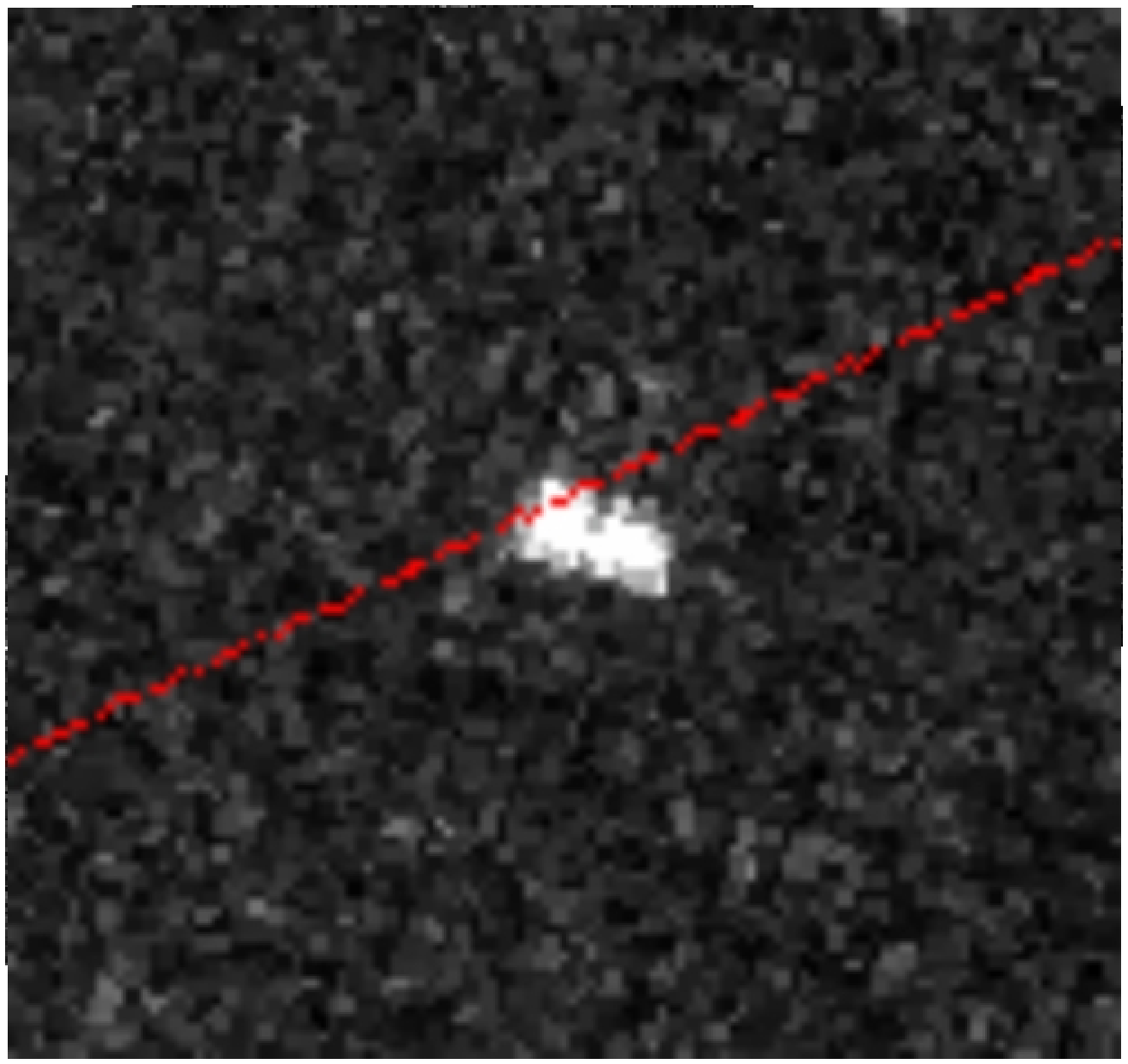} \\
\textit{a} & \textit{b} & \textit{c}
\end{tabular}
\end{center}
\caption{\textit{Panel a}: DSS-I image of a high-proper-motion star. Epoch: 1954.89. \textit{Panel b}: Combined three-colour image of the same star from DSS-II. Mean epoch: 1995.41. \textit{Panel c}: The same star on an I band NCCS image. Epoch: 2010.02. All the images are aligned with each other 
   \label{Fig:PPM}}
\end{figure}

Comparing the star coordinates from the NCCS with those of DSS showed that the star is a high-proper motion object. Fig.\ref{Fig:PPM} shows images of the star from different epochs. We compared the NCCS coordinates of the star with those measured on historical photographic surveys and estimated its proper motion to be ($\Delta\alpha$, $\Delta\delta$) = (+14$\pm$8, -56$\pm$6) mas/yr, consistent with the estimates from NOMAD (\citealt{ZAC04}): ($\Delta\alpha$, $\Delta\delta$) = (+34$\pm$22, -60$\pm$1) mas/yr. 

The star is too faint to be included in the PPM and UCAC catalogs. From its colour, distance, extinction and general astrophysical considerations we conclude tentatively that this star is probably a white dwarf located $\lesssim200$ pc from the Sun. For a final determination of its nature spectroscopic data is required.

Another example of exotic objects to be mined from the NCCS is QSOs and AGNs. We downloaded the entire \cite{VER10} catalog of quasars and AGN in the survey area, and one of the objects (out of 60) was found in the selected area. The object is located at $(\alpha,~\delta)=$ $(05^{h}36^{m}07^{s}.2$, $82^{\circ}23'14''.5)$, which is a region of low galactic extinction: E$_{B-V}$ = 0$^m$.06 (\citealt{SCH98}), and it is relatively bright: R = 15$^m$.12, I = 14$^m$.71, NUV = 18$^m$.64, FUV = 19$^m$.42. The object was confirmed spectroscopically as an AGN at a redshift z = 0.05 (\citealt{XU01}). The object was identified as an extended source by NCCS pipeline.

To demonstrate our capability to distinguish between AGNs and hot stellar objects we simulated colour-colour diagrams. AGN colours were simulated using the modified composite quasar spectrum from \cite{VAN01}. Stellar colours were simulated using the \textbf{Bruzual-Persson-Gunn-Stryker Atlas}, which is an extension of \cite{GUN83} optical stellar spectra atlas, the \textbf{\cite{PIC98} Stellar Spectral Flux Library} and the \textbf{CALSPEC} (\citealt{BOH03}) atlas, which is used for the calibration of the HST instruments. These three atlases provide a sample of $\sim$ 250 stellar spectra encompassing all spectral types and luminosity classes from super-giants to brown dwarfs and covering also a wide range of metallicity. 

The AGN spectrum was redshifted from restframe to z$\leq$3 and corected for the absorption by Lyman systems (Ly-$\alpha$ forest, Lyman limit and damped Ly-$\alpha$ systems) using the transmission function calculated from \citealt{MOL90}. AGN and stellar spectra were convolved with the response functions of the FUV, NUV, g', R and I filters to derive object flux. The magnitudes were calculated relative to Vega. 

We conclude, that optical colours alone do not provide a good AGN-stellar separation. The use of NUV-based and FUV-based colours provides an ideal separation for all the redshifts 0$<$z$<$3 tested here, though practically, only very few objects are detected in FUV by GALEX, as mentioned above. However, the effective AGN colour selection can be performed using the NUV-alone-based colour-colour diagram at 0$<$z$<$2.

We estimated the AGN yield of the project using the four following methods:
\begin{enumerate}
\item \cite{BIA05} matched GALEX sources detected by the All-sky Imaging Survey (AIS) with SDSS sources over an area of 92 deg$^2$. They selected 222 AGN candidates at redshifts $<$ 2 using their Sloan and GALEX colours. Figure 2 of \cite{BIA05} shows the distribution of the AGN candidates as a function of the Sloan $r$ magnitude. Assuming that the Johnson R magnitude is similar to the Sloan $r$ magnitude and using Figure 2 in \cite{BIA05}, we estimate the number of AGN candidates brighter than our limiting magnitude R $\leq20.2$ mag as $\sim$204 candidates. However, this number should be corrected for the area of the survey: $\frac{204\times313.6}{92}\approx695$. The estimation of \textbf{695} candidates should be also corrected for the different galactic latitude, since the SDSS was performed mostly in high galactic latitudes, while the NCCS is an intermediate-latitude survey. Therefore, the NCCS would detect less AGN candidates per total number of sources than the SDSS, as shown in Figure 4 of \cite{RIC09}. However, it is not clear from \cite{BIA05} where their matching area is located.
\item \cite{RIC09} analyzed the catalog of 1,172,157 AGN candidates selected from the SDSS by colours. Figure 18 of \cite{RIC09} shows the distribution of candidate counts per solid angle per magnitude as a function of the Sloan $i$ magnitude for the redshifts $0.3<z<2.2$. Assuming that the Johnson I magnitude is similar to the Sloan $i$ magnitude, we estimate the density of AGN candidates brighter than our limiting magnitude I $\leq19.1$ mag as $\sim$1.8 counts/deg$^2$. Thus the NCCS area of $\sim$313.6 deg$^2$ implies the estimation of $\sim$\textbf{564} candidates. However, this number is (1) not corrected for the survey latitude similar to the previous estimation and (2) is valid for a slightly different redshift range, than that of the NCCS AGN candidate selection. 
\item Figure 2 of \cite{RIC09} demonstrates the distribution of the SDSS AGN candidate counts as a function of the Sloan $i$ magnitude. The number of AGN candidates brighter than our limiting magnitude I $\leq19.1$ mag can be determined from Figure 2 as $\sim$23000 counts. However, this number should be corrected for the survey area (8417 deg$^2$ - SDSS, 313.6 deg$^2$ - NCCS): $\frac{23000\times313.6}{8417}\approx857$.The estimation of 857 candidates is (1) not corrected for the survey latitude and (2) is contaminated by higher redshift AGNs. However, Figure 18 of \cite{RIC09} shows that the contamination of the sample by higher redshift AGNs is $\lesssim$ 20\%, yielding the final estimation of $\frac{4}{5}\times857\approx$ \textbf{686} candidates. 
\item Figure 4 of \cite{RIC09} shows the ratio of AGN candidates to all sources as a function of galactic latitude \textit{b}. The ratio for the NCCS location ($17^{\circ}<b<37^{\circ}$) can be estimated from it as $\sim$1.6\%. Assuming that the NCCS is supposed to contain $\gtrsim$1,500,000 distinct sources, the number of AGN candidates is defined as: $1,500,000\times1.6\%\approx24000$ counts. However, this number should be corrected for the survey limiting magnitude. The correction factor was estimated using the above method: $\sim$23000 from the total 1,172,157 SDSS AGN candidates are brighter than I $\leq19.1$ mag. Thus the final estimation is $\frac{23000}{1,172,157}\times24000\approx$ \textbf{471} candidates.
\end{enumerate}
We estimated the NCCS AGN yield using four almost independent methods (the fourth method is partially dependent on the estimation of the third one), each one of which is biased differently. However, all the estimation methods produced remarkably similar results, allowing to calculate the average NCCS AGN yield $\sim$\textbf{600}$\pm$\textbf{100} candidates.
\vspace{-8pt}
\section{Conclusions and Further Work}
\vspace{-8pt}
We have shown that there is valuable new science that can be derived from the NCCS data set, apart from improving the global knowledge about the sky at high declinations. It is possible, therefore, to consider possible extensions of this rather limited project, as defined above.

We can relatively easily extend the surveyed region for five more degrees of declination, increasing the surveyed region to more than 700 deg$^2$. This will increase the overlap with SDSS-surveyed regions, allowing an improved cross-calibration between the two surveys. Alternatively, we could repeat the observations of the already surveyed region in one of the R and I bands to check the variability of the detected sources. 

We could merge the NCCS data with 2MASS extending the wavelength base to IR. Though 2MASS is relatively shallow, it could provide additional information about NCCS objects brighter than $\sim$16$^m$--17$^m$. Spectroscopic follow-up of the promising NCCS sources could be performed at the Wise Observatory for objects brighter than $\sim$16$^m$ or at a larger telescope for the faint ones.
\vspace{-8pt}

\acknowledgments Based on observations made with the NASA Galaxy Evolution Explorer.
GALEX is operated for NASA by the California Institute of Technology under NASA contract NAS5-98034.

This study makes use of data from the SDSS\\
(http://www.sdss.org/collaboration/credits.html).

\end{document}